# Continuous-wave Cascaded-Harmonic Generation and Multi-Photon Raman Lasing in Lithium Niobate Whispering-Gallery Resonators

Jeremy Moore, Matthew Tomes, Tal Carmon, and Mona Jarrahi
*Department of Electrical Engineering and Computer Science*
*University of Michigan, Ann Arbor, MI, 48109, USA*

We report experimental demonstration of continuous-wave cascaded-harmonic generation and Raman lasing in a millimeter-scale lithium niobate whispering-gallery resonator pumped at a telecommunication-compatible infrared wavelength. Intensity enhancement through multiple recirculations in the whispering-gallery resonator and quasi phase-matching through a nonuniform crystal poling enable simultaneous cascaded-harmonic generation up to the fourth-harmonic accompanied by stimulated Raman, two-photon, three-photon, and four-photon Raman scattering corresponding the molecular vibrational wavenumbers 632 cm$^{-1}$ and 255 cm$^{-1}$ in z-cut lithium niobate at pump power levels as low as 200mW. We demonstrate simultaneous cascaded-harmonic generation and Raman lasing by observing the spectrum of the scattered light from the resonator and by capturing the image of the decoupled light from the resonator on a color CCD camera.
© 2011 American Institute of Physics

Wavelength-conversion through various nonlinear processes is an effective approach to extend the emission wavelength of laser sources to wavelengths that are difficult to achieve through standard solid-state laser sources and are in demand for various spectroscopy and microscopy applications. The high pump power levels required for generating sufficient nonlinear optical response are generally achieved by ultra-short pump pulses. This limits the effectiveness of a number of applications such as imaging crystals and molecular states with sub-atomic-scale resolution[1,2], which will benefit from a continuous-in-time source.

Whispering-gallery resonators enable strong enhancement of various nonlinear phenomena because of their capability to confine light energy for long periods of time within small volumes[3-5]. Thus, significant reduction in the required pump power level for parametric oscillations[6-8], second[9], third[10-14], and fourth-harmonic generation[15], Raman-lasing[16-19], Erbium-lasing[20], Brillouin-lasing[21-24], and optomechanical vibrations[21-25], have been reported in high-quality-factor whispering-gallery resonators.

In this work, we experimentally demonstrate continuous-wave cascaded-harmonic generation up to the fourth-harmonic accompanied by multi-photon Raman scattering from a lithium niobate whispering-gallery resonator pumped at a telecommunication-compatible infrared wavelength and at pump power levels as low as 200mW. Our results could benefit high-harmonic and Raman studies from pulsed to continuous-wave. Quasi phase-matching to conserve momentum for the high-harmonics is achieved by using a non-uniform poling-period[26,27] and facilitated by a combination of high order transverse resonance modes[10]. Lithium niobate poling is made in stripes such that the azimuthally propagating light sees different periods as it circulates around the resonator circumference and the quasi phase matching requirements are satisfied.

A schematic of the experimental setup is shown in Fig. 1a. The whispering-gallery resonator was fabricated from a periodically poled z-cut lithium niobate substrate. A 3mm disk was cut from the wafer and the edge was mechanically polished[9] to a spherical profile. The infrared pump beam is evanescently coupled to the millimeter-scale lithium niobate whispering-gallery resonator via a diamond prism[28]. The highest whispering gallery resonator quality factor is measured to be $2\times10^7$ in the 1550nm pump wavelength range (Fig. 1b), implying up to 400-times enhancement in a circulating pump power inside the 3mm-diameter whispering-gallery resonator.

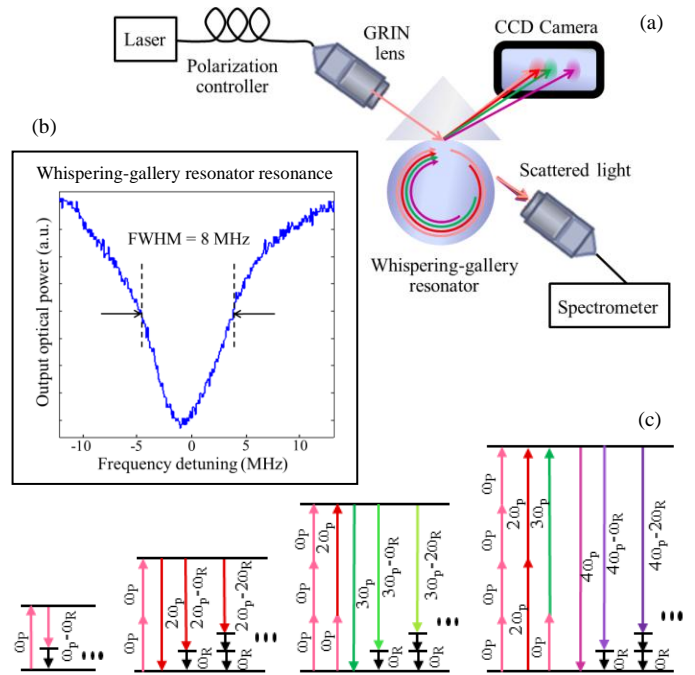

Fig. 1. a) Experimental setup for demonstrating cascaded harmonic generation and Raman lasing in the lithium niobate whispering-gallery resonator. b) Whispering-gallery resonator resonance, implying a quality factor of $2\times10^7$ at 1550nm. c) Energy-level diagrams illustrating various nonlinear processes involved in the demonstrated pump wavelength extension, including the cascaded harmonic generation, Raman scattering, cascaded Raman scattering and/or multi-photon hyper-Raman scattering in the lithium niobate whispering-gallery resonator ($\hbar\omega_P$ and $\hbar\omega_R$ represent the pump photon and Raman phonon energy levels, respectively).

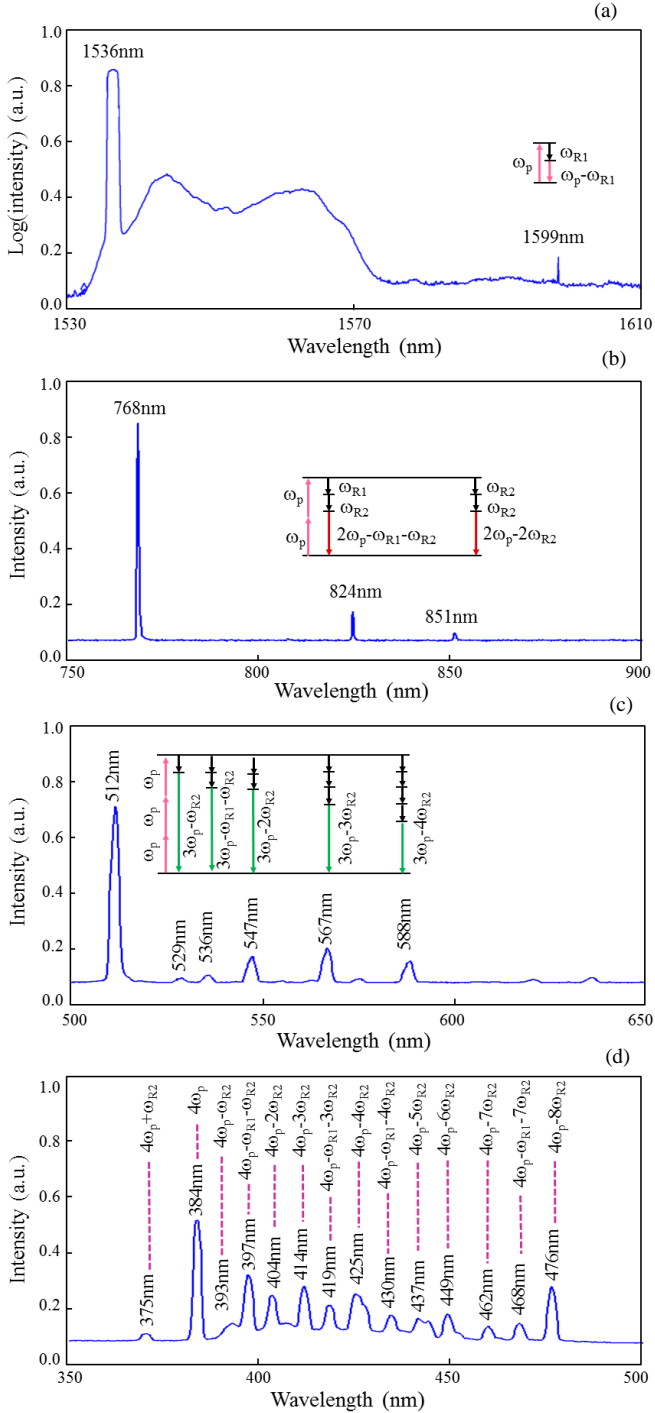

Fig. 2. Measured emission spectrum of the whispering-gallery resonator at the pump wavelength of 1536nm and pump power of 200mW at a) infrared, b) near-infrared, c) visible, and d) ultraviolet spectral ranges. The spectral lines indicate generation of the $2^{nd}$, $3^{rd}$, and $4^{th}$ harmonics as well as the stimulated Raman, two-photon, three-photon, and four-photon Raman scattering corresponding the molecular vibrational wavenumbers 632 cm$^{-1}$ and 255 cm$^{-1}$ in z-cut lithium niobate

The scattered light from the resonator is collected by a multimode optical fiber and analyzed by three spectrum analyzers which cover the infrared to ultraviolet band. The wavelength of the spectral lines of the measured spectrum and their relative intensity suggest that they are generated through cascaded harmonic generation, Raman scattering and multi-photon Raman scattering processes as shown in Fig. 1c. Within the same experimental setup, emitted light from the prism coupler is observed on a color CCD camera and analyzed for further justification of the nature of the underlying nonlinear processes.

The emission spectrum of the whispering-gallery resonator at the pump wavelength of 1536 nm and pump power of 200 mW is shown in Fig. 2. The measured spectrum indicates $2^{nd}$ (Fig. 2b), $3^{rd}$ (Fig. 2c) and $4^{th}$ (Fig. 2d) harmonic generation at 768 nm, 512 nm, and 384 nm respectively. Moreover, all three harmonics display wide tunability within the pump wavelength range of 1535-1545nm, and the $n^{th}$ harmonic wavelengths are observed to track the expected values of (*pump wavelength*)/*n*.

Figure 2a also shows a stimulated Raman line at 1599 nm which corresponds to the molecular vibrational wavenumber of 255 cm$^{-1}$ in z-cut lithium niobate crystal[29-31]. A second stimulated Raman line corresponds to the other molecular vibrational wavenumber of z-cut lithium niobate crystal at 632 cm$^{-1}$ and is not directly observed due to the limitation of the employed infrared spectrum analyzer. However, the proof of excitation of this Raman line is evident through several multi-photon Raman lines observed in the near-infrared (Fig. 2b), visible (Fig. 2c), and ultraviolet (Fig. 2d) wavelength ranges. More specifically, for a pump photon energy of $\hbar\omega_P$, the observed mid-infrared spectral lines at 824nm and 851nm match with the two-photon Raman lines $2\omega_p-\omega_{R1}-\omega_{R2}$ and $2\omega_p-2\omega_{R2}$, where $\hbar\omega_{R1}$ and $\hbar\omega_{R2}$ are the phonon energy levels of the 255 cm$^{-1}$ and 632 cm$^{-1}$ Raman lines, respectively. The observed visible spectral lines at 529nm, 536nm, 547nm, 567nm, and 588nm match with the three-photon Raman lines $3\omega_p-\omega_{R2}$, $3\omega_p-\omega_{R1}-\omega_{R2}$, $3\omega_p-3\omega_{R2}$, and $3\omega_p-4\omega_{R2}$. The observed ultraviolet spectral lines at 375nm, 393nm, 397nm, 404nm, 414nm, 419nm, 425nm, 430nm, 437nm, 449nm, 462nm, 468nm, and 476nm match with the four-photon Raman lines $4\omega_p+\omega_{R2}$, $4\omega_p-\omega_{R2}$, $4\omega_p-\omega_{R1}-\omega_{R2}$, $4\omega_p-2\omega_{R2}$, $4\omega_p-3\omega_{R2}$, $4\omega_p-\omega_{R1}-3\omega_{R2}$, $4\omega_p-4\omega_{R2}$, $4\omega_p-\omega_{R1}-4\omega_{R2}$, $4\omega_p-5\omega_{R2}$, $4\omega_p-6\omega_{R2}$, $4\omega_p-7\omega_{R2}$, $4\omega_p-\omega_{R1}-7\omega_{R2}$, and $4\omega_p-8\omega_{R2}$. We believe that both hyper-Raman[19] and cascaded Raman processes[16] are involved in the observed multi-photon Raman scattering that extends the spectrum of the pump and the $2^{nd}$, $3^{rd}$, and $4^{th}$ harmonics to a train of lines that are separated by $\omega_{R1}$ and $\omega_{R2}$.

Further evidence for the nature of the observed nonlinear processes is achieved by monitoring the whispering-gallery resonator out-coupled light on a color CCD camera (Fig. 3). Spectral filters are employed to prevent saturation of the camera by the $2^{nd}$ and $3^{rd}$ harmonics, and the ultraviolet light is observed by coating the CCD with a UV fluorescent ink. The spatial distribution of the beams observed on the CCD correspond to the expected positions of the $2^{nd}$, $3^{rd}$, and $4^{th}$ harmonic wavelengths (769 nm, 512.7nm, and 384.5 nm, respectively for a pump wavelength of 1538 nm) at a pump power of 200mW. The spatial separation of the near-infrared, visible and ultraviolet light is due to dispersion in the prism, as well as chromatic and waveguide-mode dispersion in the resonator. Apart from the cascaded harmonic spots, whenever a Raman line was seen on the spectrum analyzer, a corresponding color (e.g. yellow, orange, and shades of green) was showing up in the proper region on the CCD picture. This serves as another verification for the Raman process.

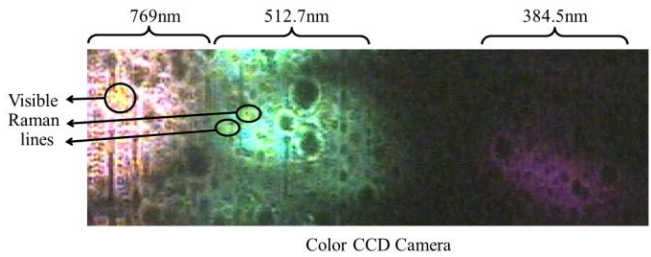

Fig. 3. Whispering-gallery resonator out-coupled light observed on a color CCD camera at a pump wavelength of 1538nm. The red, green and purple spots correspond to the 1$^{st}$, 2$^{nd}$, and 3$^{rd}$ harmonics. The less bright yellow and orange spots correspond to the three-photon Raman lines, yellow-shifted with respect to the third harmonic.

In conclusion, we experimentally demonstrate continuous-wave cascaded-harmonic generation up to the fourth-harmonic accompanied by stimulated Raman, two-photon, three-photon, and four-photon Raman scattering corresponding the molecular vibrational wavenumbers 632 cm$^{-1}$ and 255 cm$^{-1}$ in z-cut lithium niobate at pump power levels as low as 200mW. By excitation of higher order harmonics one can push the cascaded-harmonic and multi-photon Raman emission to deep-ultraviolet. However, such demonstration is not possible for our current device due to the high absorption of deep-ultraviolet light in lithium niobate. The continuous progress in developing ultra-high-quality-factor resonators suggests the possibility of extending the continuous-wave high-harmonic generation demonstrated here to over 4-octaves of frequency band. In this regard, high-quality-factor whispering-gallery resonators made of crystals with $\chi^{(3)}$ effect, such as CaF$_2$, are very promising due to their broadband transparency window (0.18 – 8 μm for CaF$_2$).

The authors would like to thank Prof. Harald Schwefel, Prof. Mani Hossein-Zadeh at the University of New Mexico, and Prof. Bahram Jalali's group at UCLA for advice and assistance with the experiment, and Opticology, Inc. for assistance with fabrication. Matthew Tomes is supported by a Graduate Research Fellowship from the National Science Foundation. This work is supported by National Science Foundation ENG-ECCS-065614, and by the Air Force Office of Scientific Research Young Investigator Award under contract number FA9550-10-1-0078.